\documentclass[
  ,final            
  ]
  {aipproc}

\layoutstyle{6x9}

\usepackage[latin1]{inputenc}
\usepackage[T1]{fontenc}
\usepackage[english]{babel}
\usepackage{epsfig}
\usepackage{amssymb}
\usepackage[centertags]{amsmath}
\usepackage{amsthm}
\usepackage{amsfonts}
\usepackage{makeidx}
\usepackage{graphics}
\usepackage{graphicx}
\usepackage{psfrag}
\usepackage{subfigure}
\usepackage{slashed}
\usepackage{dsfont}
\usepackage{eucal}
\usepackage{eufrak}
\usepackage{color}
\usepackage{bm}
\usepackage{fancybox}
\usepackage{footnpag}
\usepackage{latexsym}
\usepackage{appendix}
\usepackage{fancyhdr}
\usepackage{textcomp}
\usepackage{latexsym}
\usepackage{slashed} 
\usepackage{fancyhdr}
\usepackage[marginal]{footmisc}
\usepackage{appendix}
\usepackage{dsfont}
\usepackage{feynmf}

\def \non {\nonumber}
\def \noi {\noindent}
\def \ra {\rightarrow}

\def \pr {\prime}

\def \mau {\geqslant}

\def \fr {\displaystyle\frac}

\def\laq{~\raise 0.4ex\hbox{$<$}\kern -0.8em\lower 0.62
ex\hbox{$\sim$}~}
\def\gaq{~\raise 0.4ex\hbox{$>$}\kern -0.7em\lower 0.62
ex\hbox{$\sim$}~}

\begin{document}

\title[]{Recent issues in open and hidden charm spectroscopy}
\date{}



\author{Stefano~Nicotri\thanks{stefano.nicotri@ba.infn.it}}{
  address={Universit\`a degli Studi di Bari and INFN Sezione di Bari~-~Italy}}



\begin{abstract}
  I present a brief review of  results obtained both in open
  and hidden charm spectroscopy, discussing the
  interpretation of $D_{sJ}(2860)$, $D_{sJ}(2700)$ and $X(3872)$.
\end{abstract}

\maketitle


\section{Introduction}

Many new states have been recently observed in the open and hidden
charm sector: $D_{sJ}^*(2317)$, $D_{sJ}(2460)$, $D_{sJ}(2700)$,
$D_{sJ}(2860)$, $D_{0}^*(2308)$, $D_{1}^\prime(2440)$, together with
$h_c$, $\eta_c^\pr$, $X(3872)$, $X(3940)$, $Y(3940)$, $Z(3930)$,
$Y(4260)$.... \cite{Reviews}. The need for theoretical
interpretation of this states comes not only from the request of
organizing the particle ``zoology'', but also from the interesting
possibility of identifying new ``exotic'' structures. This is what
we would like to briefly discuss in the next section devoted to the
$c\bar s$ sector, with particular attention to the $D_{sJ}(2860)$
and a few words on $D_{sJ}(2700)$, and in the third section devoted
to the hidden charm sector and, in particular, to the interpretation
of $X(3872)$.

\section{Charmed-strange mesons and $D_{sj}(2860)$}

The classification of the $c\bar s$ states is easier in the
heavy-quark limit $m_c\ra\infty$. In this limit the spin $s_Q$ of
the heavy quark  and  the angular momentum $s_\ell$ of the light
degrees of freedom: $s_\ell=s_{\bar q}+ \ell$ ($s_{\bar q}$ light
antiquark spin, $\ell$  orbital angular momentum of the light
degrees of freedom relative to the heavy quark) are decoupled, and
the spin-parity  $s_\ell^P$  is conserved in strong interactions
\cite{HQET}. This makes possible to classify mesons into doublets
labeled by $s_\ell^P$ (where $P$ is the parity), each containing a
couple of meson of spin-parity $J^P=(s_\ell^P-1/2,s_\ell^P+1/2)$ and
degenerate in mass. Mass differences between members of the same
doublet are of order ${\cal O}(1/m_c)$. The standard classification
of known $c\bar s$ states in this scheme is given in
Table~\ref{organiz} \cite{PDG}. The states labeled by
$P^{*\prime}_{s2}$ and $P^{*}_{s2}$ are still to be discovered; we
discuss here a possible identification of $P_{s3}$ and
$P^{*\prime}_{s1}$.

\begin{table}[ht]
\caption{$c \bar s$ states organized according to $s_\ell^P$ and
$J^P$. The mass of  known mesons    is indicated. } \label{doublets}
\label{organiz}
\begin{tabular}{c|cccccc}
\hline $s_\ell^P  $ & ${1\over 2}^-$ & ${1\over 2}^+$ &
${3\over2}^+$ & ${3\over 2}^-$ & ${5\over 2}^-$ \\ \hline
$J^P=s_\ell^P-{1\over 2}$ & $D_s (1965) \,\, (0^-)  $ & $D^*_{sJ}(2317) \,\, (0^+)$ & $D_{s1}(2536) \,\, (1^+) $ & $(P^{*\prime}_{s1})  \,\, (1^-)$ & $(P^{*\prime}_{s2}) \,\,(2^-)$\\
$J^P=s_\ell^P+{1\over 2}$ & $D_s^*(2112) \,\, (1^-)$ & $D_{sJ}(2460)
\,\, (1^+)$  & $D_{s2}(2573) \,\,
(2^+)  $ & $(P^{*}_{s2}) \,\, (2^-)$ & $ (P_{s3}) \,\, (3^-)$\\
\hline
\end{tabular}
\end{table}

\noi In the above classification the $D_{sJ}(2860)$, observed by
BaBar in the $DK$ invariant mass distribution \cite{Aubert:2006mh},
can be either a $J^P=1^-$ $ s_\ell^P={3 \over 2}^-$ state, or a
$J^P=3^-$   $ s_\ell^P={5 \over 2}^-$ state, the $P_{s3}$, i.e. a
state with $\ell=2$ and lowest radial quantum number. Another
possibility is that $D_{sJ}(2860)$ is a radial excitation of the
$J^P=1^-$   $ s_\ell^P={1 \over 2}^-$ state ($D_s^{*\prime}$), of
the  $J^P=0^+$ $ s_\ell^P={1 \over 2}^+$ state (first radial
excitation of $D_{sJ}^*(2317)$) or of the $J^P=2^+$ $ s_\ell^P={3
\over 2}^+$ state ($D_{s2}^\prime$). The $J^P$ assignment  can be
done considering the decay modes and width. In order to evaluate
them we define the fields representing the various doublets: $H_a$
for $s_\ell^P={1\over2}^-$, $S_a$ and $T_a$ for
$s_\ell^P={1\over2}^+$ and $s_\ell^P={3\over2}^+$, respectively, and
$X_a$ and $X^\prime_a$ for the doublets corresponding  to $\ell=2$,
$s_\ell^P={3\over2}^-$ and $s_\ell^P={5\over2}^-$, respectively:
\begin{eqnarray}
H_a & =&
\frac{1+{\rlap{v}/}}{2}[P_{a\mu}^*\gamma^{\,\mu}-P_a\gamma_5]
\label{neg}  \hspace*{1cm} ,  \hspace*{1cm}
S_a = \frac{1+{\rlap{v}/}}{2} \left[P_{1a}^{\prime \mu}\gamma_\mu\gamma_{\,5}-P_{0a}^*\right]  \,\,\,, \non \\
T_a^\mu &=&\frac{1+{\rlap{v}/}}{2} \left\{ P^{\,\mu\nu}_{2a}
\gamma_\nu -P_{1a\nu} \sqrt{3 \over 2} \gamma_{\,5} \left[ g^{\mu
\nu}-{1 \over 3} \gamma^{\,\nu} (\gamma^{\,\mu}-v^\mu) \right]
\right\}  \,\,\, , \non \\
X_a^\mu &=&\frac{1+{\rlap{v}/}}{2} \left\{ P^{*\mu\nu}_{2a}
\gamma_{\,5} \gamma_\nu -P^{*\prime}_{1a\nu} \sqrt{3 \over 2} \left[
g^{\mu \nu}-{1 \over 3} \gamma^{\,\nu} (\gamma^{\,\mu}-v^\mu)
\right]
\right\}   \,\,\,\, ,  \label{pos2}  \\
X_a^{\prime \mu\nu} &=&\frac{1+{\rlap{v}/}}{2} \left\{
P^{\,\mu\nu\sigma}_{3a} \gamma_\sigma -P^{*'\alpha\beta}_{2a}
\sqrt{5 \over 3} \gamma_{\,5} \left[ g^\mu_\alpha g^\nu_\beta -{1
\over 5} \gamma_\alpha g^\nu_\beta (\gamma^{\,\mu}-v^\mu)-  {1 \over
5} \gamma_\beta g^\mu_\alpha (\gamma^{\,\nu}-v^\nu) \right] \right\}
\non
\end{eqnarray}

\noi with the various operators annihilating mesons of four-velocity
$v$ which is conserved in  strong interaction. The interaction of
these particles with the octet of light pseudoscalar mesons,
represented by $\displaystyle \xi=e^{i {\CMcal M}/f_\pi}$,
$\Sigma=\xi^2$ and  the matrix ${\CMcal M}$ containing $\pi, K$ and
$\eta$ fields:
\begin{equation}
{\CMcal M}= \left(\begin{array}{ccc}
\sqrt{\frac{1}{2}}\pi^0+\sqrt{\frac{1}{6}}\eta & \pi^+ & K^+\nonumber\\
\pi^- & -\sqrt{\frac{1}{2}}\pi^0+\sqrt{\frac{1}{6}}\eta & K^0\\
K^- & {\bar K}^0 &-\sqrt{\frac{2}{3}}\eta
\end{array}\right)
\end{equation}
($f_{\pi}=132 \; $ MeV) can be described by the interaction
lagrangians:
\begin{eqnarray}
{\cal L}_H &=& \,  g \, Tr [{\bar H}_a H_b \gamma_\mu \gamma_5 {\cal
A}_{ba}^\mu ] \non \\
{\cal L}_S &=& \,  h \, Tr [{\bar H}_a S_b \gamma_\mu \gamma_5 {\cal
A}_{ba}^\mu ]\, + \, h.c. \,, \non \\
{\cal L}_T &=&  {h^\prime \over \Lambda_\chi}Tr[{\bar H}_a T^\mu_b
(i D_\mu\slashed{\cal A}+i\slashed{D} { \cal A}_\mu)_{ba} \gamma_5
] + h.c.   \label{lag-hprimo}   \\
{\cal L}_X & = & \fr{k^\prime}{\Lambda_\chi} Tr[{\bar H}_a X^\mu_b
(i D_\mu\slashed{\cal A}+i\slashed{D}{\cal A}_\mu)_{ba} \gamma_5
] + h.c.  \non \\
{\cal L}_{X^\prime} &=&  {1 \over {\Lambda_\chi}^2}Tr[{\bar H}_a
X^{\prime \mu \nu}_b [k_1 \{D_\mu, D_\nu\} {\cal A}_\lambda+ k_2
(D_\mu D_\nu { \cal A}_\lambda + D_\nu D_\lambda { \cal
A}_\mu)]_{ba}  \gamma^\lambda \gamma_5] + h.c.  \non
\end{eqnarray}
where $\Lambda_\chi$ is  the chiral symmetry-breaking scale (
$\Lambda_\chi = 1 \, $ GeV). ${\cal L}_S$ and ${\cal L}_T$ describe
transitions of positive parity heavy mesons with the emission of
light mesons in $s-$ and $d-$ wave, respectively, $g, h$ and
$h^\prime$ being  effective coupling constants, while ${\cal L}_X$
and ${\cal L}_{X^\prime}$  describe the transitions of higher mass
mesons of negative parity with the emission of light mesons in $p-$
and $f-$ wave with coupling constants $k^\prime$, $k_1$ and $k_2$.

In Table \ref{ratios} the ratios $ {\Gamma( D_{sJ} (2860) \to D^*K)
\over \Gamma( D_{sJ} (2860)\to DK) }$ and $ {\Gamma( D_{sJ}
(2860)\to D_s \eta) \over \Gamma( D_{sJ} (2860)\to DK)  }$ obtained
in this framework for various quantum number assignments to $D_{sJ}
(2860)$ \cite{Colangelo:2006rq} are shown.
\begin{table}[ht]
    \caption{Predicted ratios
    $\displaystyle {\Gamma( D_{sJ} \to D^*K) \over \Gamma( D_{sJ} \to DK)
 }$   and   $\displaystyle {\Gamma( D_{sJ} \to D_s \eta) \over \Gamma( D_{sJ} \to DK)  }$ (with $DK=D^0K^+ + D^+ K_S^0$) for  various assignment
 of quantum numbers to  $D_{sJ}(2860)$.}
    \label{ratios}
    \begin{tabular}{| c | c | c | c |}
      \hline
 $D_{sJ}(2860) $  &  $D_{sJ}(2860) \to DK $&$\displaystyle{\Gamma( D_{sJ} \to D^*K) \over \Gamma( D_{sJ} \to DK)
 }$  &   $\displaystyle{\Gamma( D_{sJ} \to D_s \eta) \over \Gamma( D_{sJ} \to DK)  }$
\\ \hline
 $s_\ell^p={1\over 2}^-$, $J^P=1^-$,  rad. excit. & $p$-wave &$1.23$& $0.27$ \\
$s_\ell^p={1\over 2}^+$, $J^P=0^+$,  \hspace*{0.3cm} " & $s$-wave &$0$& $0.34$ \\
$s_\ell^p={3\over 2}^+$, $J^P=2^+$,  \hspace*{0.3cm}  " & $d$-wave &$0.63$& $0.19$\\
$s_\ell^p={3\over 2}^-$, $J^P=1^-$     \hspace*{0.6cm} & $p$-wave  & $0.06$& $0.23$ \\
$s_\ell^p={5\over 2}^-$, $J^P=3^-$     \hspace*{0.6cm}& $f$-wave  & $0.39$& $0.13$ \\
    \hline
    \end{tabular}
\end{table}
These ratios   can be used to exclude some assignments. Indeed,
since a $D^*K$ signal has not been observed in the $D_{sJ}(2860)$
mass range, the production of $D^* K$ is not favoured and therefore
$D_{sJ}(2860)$ is not a radial excitation of $D_s^*$ or $D_{s2}$.
The assignment $s_\ell^p={3\over 2}^-$, $J^P=1^-$ can also be
excluded:   the  width  $\displaystyle \Gamma(D_{sJ}(2860)\to DK) $
obtained using (\ref{lag-hprimo}) would be too big using $k^\prime
\simeq h^\prime\simeq 0.45\pm0.05$ \cite{Colangelo:2005gb}, and
there is no reason to presume that the coupling constant $k^\prime$
is sensibly smaller.

\noi In the case of the assignment  $s_\ell^p={1\over 2}^+$,
$J^P=0^+$, proposed in \cite{vanBeveren:2006st}, the decay
$D_{sJ}(2860)\to D^*K$ is forbidden and  the transition into $DK$
occurs in $s-$wave. The coupling costant for  the lowest radial
quantum number  is  $h\simeq-0.55$ \cite{Colangelo:1995ph}; using
this value for $\tilde h$ we would obtain  $\Gamma(D_{sJ}(2860)\to
DK)\mau1$ GeV.  It is reasonable to suppose that  $|\tilde h| <
|h|$,  although no information is available about  couplings of
radially excited heavy-light mesons to low-lying states: the
experimental width corresponds to  $\tilde h=0.1$.  A large signal
in the $D_s \eta$ channel would also be expected. A problem is that,
if $D_{sJ}(2860)$ is a $0^+$ radial excitation, its  partner with
$J^P=1^+$  would decay to $D^* K$ with a width of the order of 40
MeV. Since   both the lowest lying states with $J^P= 0^+$ and $1^+$,
$D^*_{sJ}(2317)$ and $D_{sJ}(2460)$,  are produced in charm
continuum at $B$ factories, one must invoke an exotic mechanism to
explain the absence of the $D^*K$ signal  at  energy around $2860$
MeV.

In the last case  $s_\ell^p={5\over 2}^-$, $J^P=3^-$ the narrow $DK$
width is due to the   kaon momentum  suppression: $\displaystyle
\Gamma(D_{sJ}(2860)\to DK)\propto  q_K^7$.  A smaller but non
negligible signal in the $D^*K$ mode is predicted,  and  a small
signal in the $D_s \eta$ mode is also expected. 
Moreover, a fact that supports this assignment is that
$D_{sJ}(2860)$ with $J^P=3^-$ is not expected to be produced in non
leptonic $B$ decays such as $B^0 \to D^- D_{sJ}(2860)^+$ and $B^+
\to  \bar D^0 D_{sJ}(2860)^+$ and indeed in the Dalitz plot analysis
of $B^+ \to \bar D^0 D^0 K^+$ Belle found no signal of
$D_{sJ}(2860)$ \cite{belle2715}.

The conclusion  of our study  is that $D_{sJ}(2860)$ is likely  a
$J^P=3^-$ state, a  predicted high mass, high spin and relatively
narrow $c \bar s$ state \cite{Colangelo:2000jq}. This conclusion is
confirmed by a recent lattice QCD analysis \cite{Koponen:2007nr}.
Its non-strange partner $D_3$, if the mass splitting
$M_{D_{sJ}(2860)}-M_{D_3}$ is of the order of the strange quark
mass, is  also expected to be narrow: $\Gamma(D_{3}^+\to D^0 \pi^+
)\simeq 37$ MeV. It  can  be produced in semileptonic and in non
leptonic $B$ decays, such as $B^0 \to D_3^- \ell^+  \bar \nu_\ell$
and $B^0 \to D_3^- \pi^+$ \cite{Colangelo:2000jq}: its observation
could  be used to confirm the quantum number assignment to the
resonance $D_{sJ}(2860)$ found by BaBar.

An analogous study for $D_{sJ}(2700)$ ($J^P=1^-$) \cite{belle2715}
discussing how to distinguish between the two possible quantum
number assignments $s_\ell^P=1/2^-,\;n=1$ or $s_\ell^P=3/2^-,\;n=0$
\cite{rizzi}, shows that the ratio $\Gamma(D_{sJ} \to D^* K) \over
\Gamma(D_{sJ} \to D K)$ is different in the two scenarios and so it
may be useful to understand the right identification. Other
investigations of $D_{sJ}(2700)$ and $D_{sJ}(2860)$ involving
potential models can be found in \cite{Zhang}.

\section{Hidden charm sector and $X(3872)$}

One of the most interesting mesons in the hidden charm sector is the
$X(3872)$, discovered in the $J/\psi \pi^+ \pi^-$ invariant mass
distribution in  $B$ decays and in $p \bar p$ collisions
\cite{Choi:2003ue}, with $M(X)=3871.2 \pm 0.5$ MeV  and
$\Gamma(X)<2.3$ MeV ($90\%$ C.L.) \cite{PDG}. The $\pi^+ \pi^-$
spectrum  is peaked  for large invariant mass \cite{pipispectrum}.
$X(3872)$  was not observed in $e^+ e^-$ annihilation and in $\gamma
\gamma$ fusion, and there is also no evidence of the existence of
charged partners. The observation of the $X \to J/\psi \gamma$ mode
\cite{belle3p} indicates that the  charge conjugation of the state
is C=+1; angular distribution studies show that the most likely
quantum number assignment is $J^{PC}=1^{++}$ \cite{Abe:2005iy}.

Since another hadronic  decay mode was observed for  $X(3872)$: $X
\to J/\psi \pi^+ \pi^- \pi^0$  with  $\frac{B(X \to J/\psi \pi^+
\pi^- \pi^0)}{B(X \to J/\psi \pi^+ \pi^- )}=1.0 \pm 0.4 \pm 0.3$
\cite{belle3p,Y3930}, there are G-parity violating $X$ transitions:
this suggested the conjecture that  $X(3872)$ is not a charmonium
$\bar c c$ state. Indeed, the coincidence between the $X$ mass  as
averaged by PDG and the $D^{*0} \overline D^0$ mass inspired the
proposal that $X(3872)$ could be a molecular quarkonium \cite{okun},
a $D^{*0}$ and $\overline D^0$ bound state with small binding energy
due to a single pion exchange \cite{molec}. Such an interpretation
would allow to account for a few properties of $X(3872)$. For
example, describing the wave function of $X(3872)$ through various
hadronic components \cite{voloshin1}:
\begin{equation}
|X(3872)>=a \, |D^{*0} \bar D^0+ \bar D^{*0}  D^0> + b \, |D^{*+}
D^-+  D^{*-}  D^+> + \dots
\end{equation}
(with $|b| \ll |a|$) one could explain why this state seems not to
have  definite isospin, why the mode $X \to J/\psi \pi^0 \pi^0$ was
not  found, and why, if the molecular binding mechanism is truly
provided by a single pion exchange (however, this is a controversial
point), there are no $D \overline D$ molecular states. Anyway,
concerning the large value of the ratio $\frac{B(X \to J/\psi \pi^+
\pi^- \pi^0)}{B(X \to J/\psi \pi^+ \pi^- )}$ one has to consider
that phase space effects in two and three pion modes are very
different. The ratio of the amplitudes is smaller: $ \frac{A(X \to
J/\psi \rho^0)}{A(X \to J/\psi \omega)}\simeq 0.2$, so that the
isospin violating amplitude is 20\% of the isospin conserving one,
an effect that could be related to the mass difference between
neutral and charged $D$ mesons, considering  the contribution of
$DD^*$ intermediate states to $X$ decays.  It has also been
suggested that the molecular interpretation would imply that the
radiative decay in neutral $D$ mesons: $X \to D^0 \bar D^0 \gamma$
should be dominant with respect to $X \to D^+ D^- \gamma$
\cite{voloshin1}. However, assuming that $X(3872)$ is an ordinary
$J^{PC}=1^{++}$ charmonium and describing the $X(3872)\to D \bar D
\gamma$ amplitude by diagrams with $D^*$ and $\psi(3770)$ as
intermediate particles, the ratio $R={\Gamma(X \to D^+ D^-
\gamma)\over \Gamma(X \to D^0 \overline D^0 \gamma)}$ is small, and
it is tiny in a wide range of the hadronic parameters governing the
decays, therefore  $R\ll 1$ is not peculiar of a molecular
quarkonium $X(3872)$, but it is mostly a phase space effect
\cite{Colangelo:2007ph}.

The photon spectrum is drawn in fig.~1 for extremal values of the
hadronic parameters governing the transition. When the intermediate
$D^*$ dominates the decay amplitude, the photon spectrum in the $D^0
\bar D^0 \gamma$ mode coincides with the line corresponding to the
$D^*$ decay at $E_\gamma \simeq 139$ MeV. The narrow peak is
different from the line shape expected in a molecular description,
being broader for larger binding energy. On the other hand, the
photon spectrum in the charged $D^+ D^- \gamma$ mode is broader,
with a peak at $E_\gamma \simeq 125$ MeV, the total $X \to D^+  D^-
\gamma$ rate being severely suppressed with respect to $X \to D^0
\bar D^0 \gamma$. Instead, in the range where $\psi(3770)$ gives the
main contribution, a peak at $E_\gamma \simeq 100$ MeV appears in
neutral and charged $D$ modes, in the first case together with the
structure at $E_\gamma \simeq 139$ MeV. This spectrum was previously
described and the radiative decay was interpreted as due to the
$\bar c c$ core of $X(3872)$\cite{voloshin1}. We then suggest that
its experimental investigation could be a better tool to shed light
on the structure of this meson.
\begin{figure}[h]
 \includegraphics[width=0.40\textwidth] {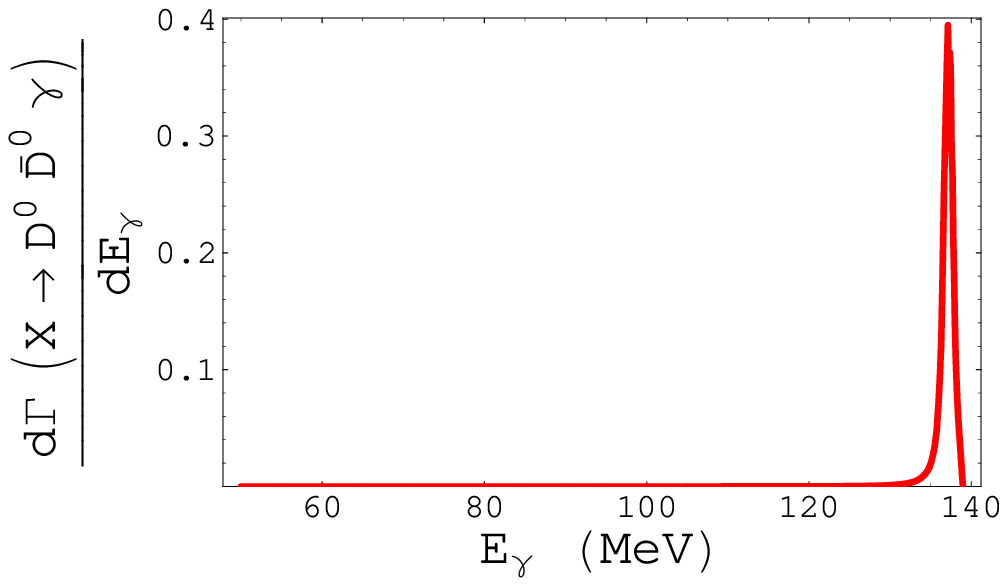} \hspace{0.25cm}
 \includegraphics[width=0.40\textwidth]{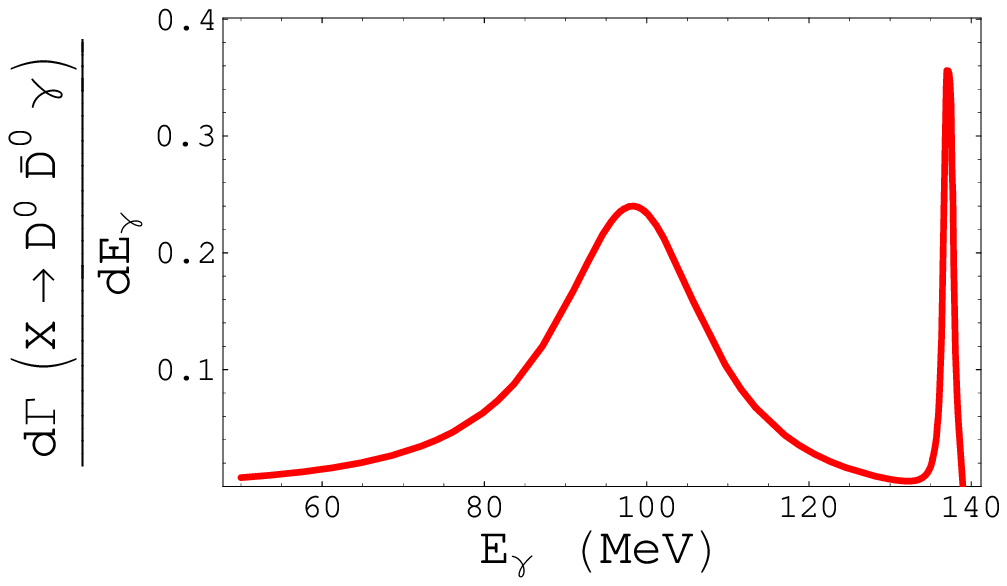}\hspace{0.25cm}
\end{figure}
\begin{figure}[h]
 \includegraphics[width=0.45\textwidth] {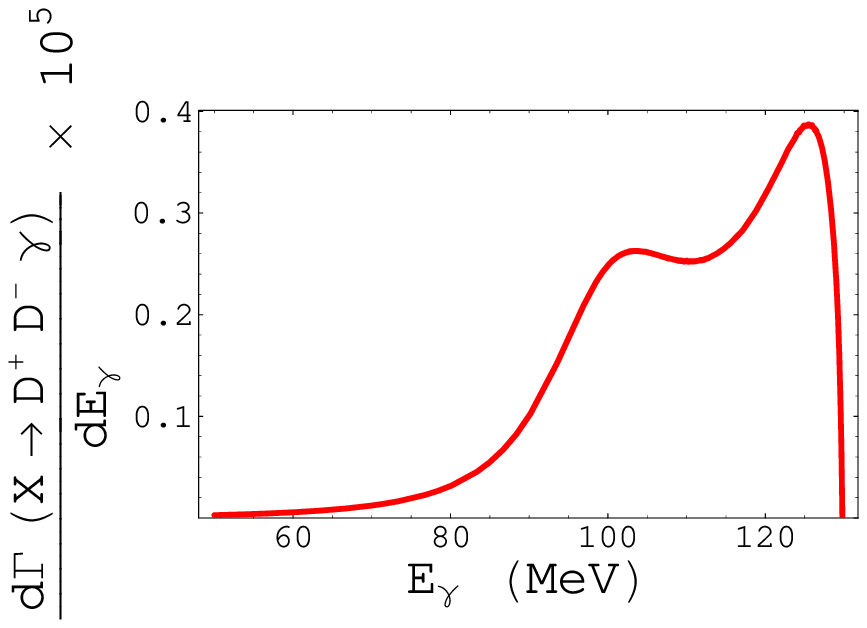} \hspace{0.25cm}
 \includegraphics[width=0.40\textwidth] {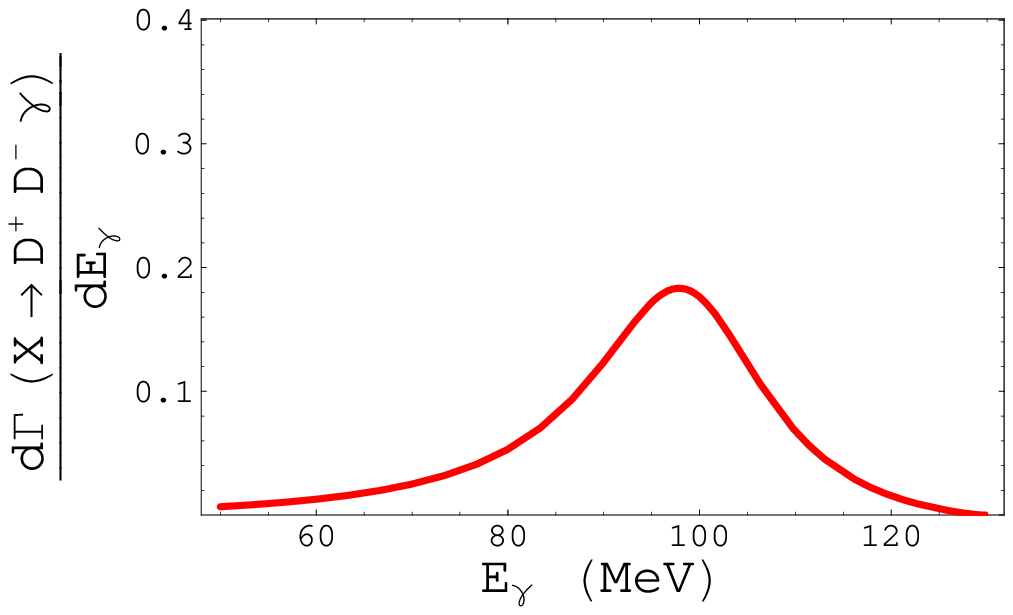}
\caption{\footnotesize{Photon spectrum (in arbitrary units) in $X
\to D^0 \bar D^0 \gamma$ (top) and  $X \to D^+ D^- \gamma$ (bottom)
decays  for values of the hadronic parameter for which the
intermediate $D^*$ dominates (left) or the intermediate $\psi(3770)$
dominates (right).}}\label{fig:spectra}
\end{figure}

\section{Conclusions}

In the open charm sector, the $c\bar s$ meson, $D_{sj}(2860)$ seems
to be a $J^P=3^-$, a member of the $s_\ell^P=5/2^-$ doublet. We have
also briefly discussed about the possible quantum number assignment
of $D_{sj}(2700)$. In both cases the analysis of the $D^*K$ mode is
crucial.

In the hidden charm sector we have described the meson $X(3872)$,
focusing our attention on its radiative decays and pointing out that
the smallness of the ratio $R={\Gamma(X \to D^+ D^- \gamma)\over
\Gamma(X \to D^0 \overline D^0 \gamma)}$ is not a smoking gun for
the molecular nature of this state. The experimental investigation
of the photon spectrum could be useful to shed more light on this
puzzling hadron.

\begin{theacknowledgments}
  I thank the workshop organizers for the nice week in Martina~Franca.
  I also acknowledge P.~Colangelo and F.~De~Fazio for collaboration.
  This work has been supported in part by the EU Contract  No. MRTN-CT-2006-035482, "FLAVIAnet".
\end{theacknowledgments}

\end{document}